\documentstyle[preprint,eqsecnum,aps]{revtex}

\begin{document}
\title{Conformal general relativity contains the quantum}
\author{R. Bonal\thanks{bonal@uclv.etecsa.cu }, I. Quiros\thanks{
israel@uclv.etecsa.cu} and R. Cardenas\thanks{rcardenas@uclv.etecsa.cu}}
\address{Dpto. Fisica. Universidad Central de Las Villas. Santa Clara CP 54830. Villa Clara. Cuba}
\date{\today}
\maketitle

\begin{abstract}
Based on the de Broglie-Bohm relativistic quantum theory of motion we show
that the conformal formulation of general relativity, being linked with a Weyl-integrable geometry, may implicitely contain the quantum effects of matter. In this context the Mach's principle is discussed.
\end{abstract}

In recent papers\cite{qbc,iq} the singularity problem
has been treated from a point of view that is based upon the conformal
transformation technique. Under a conformal transfomation of the spacetime metric

$$
\hat g_{ab}=\Omega^{2}(x)g_{ab},  
\eqno{(1)}
$$
where $\Omega(x)$ is a non-vanishing smooth function, the
spacetime coincidences [coordinates] are unchanged. Correspondingly, the
experimental observations [measurements], being nothing but just
verifications of these coincidences are unchanged too by this conformal
transformation.

Dicke showed that the transformation (1) can be interpreted as a
point-dependent transformation of the units of length, time and mass \cite
{dk}. In fact, if one takes the arc-length as one's unit of measurement of
length and time [the speed of ligth $c=1$] hence, since $d\hat s=\Omega(x)
ds$ , the unit of length in the conformal frame is point-dependent even if $ds$ is a constant along a geodesic in the original frame. In the same way, in Brans-Dicke (BD) theory \cite{bdk} the unit of mass transforms under (1) as $\hat m=\Omega(x)^{-1}m$ . Hence, even if the unit of mass in the original frame $m$ is a constant, in the conformal frame $\hat m$ will be point-dependent.

In Ref.\cite{iq} the requirement that the laws of physics must be
invariant not only under general coordinate transformations but, also, under point-dependent transformations of the units of lenght, time and mass, was raised to a cathegory of a postulate and was used there as a selection principle while studying diferent generic effective theories of gravity. It has been shown there that the only surviving candidate for a final low-energy theory of spacetime [among those studied therein] is the one with non-minimal coupling of the dilaton both to curvature and to the Lagrangian of the matter fields. The requirement of invariance under point-dependent transformations of units leads one to consider geometries that admit units of measure that may vary along transport\cite{iq}. These geometries are called generically as Weyl geometries\cite{weyl}. 

Usually the effective theories with minimal coupling of the matter fields, i. e., with the matter part of the action of the kind

$$
S_{matter}=16\pi\int d^4x\sqrt{-g}\;L_{matter},
\eqno{(2)}
$$
are linked with manifolds of Riemannian structure since, in this case, the [time-like] matter particles follow free-motion paths that are solutions of the differential equation 

$$
\frac{d^2 x^a}{ds^2}+\{^{\;\;a}_{mn}\}\frac{dx^m}{ds}\frac{dx^n}{ds}=0,
\eqno{(3)}
$$
where $\{^{\;a}_{bc}\}\equiv\frac{1}{2}g^{an}(g_{bn,c}+g_{cn,b}-g_{bc,n})$ are the Christoffel symbols of the metric $g_{ab}$. I. e., these coincide with the Riemannian geodesics of the metric. In general a Riemann configuration is characterized  by the requirement that the covariant derivatives of the metric tensor vanish, i.e.,

$$
g_{ab;c}=0,
\eqno{(4)}
$$
where semicolon denotes covariant differentiation in a general affine sense.\footnotemark\footnotetext{We use, mainly, the notation of Ref.\cite{novello}} Fulfillment of this condition leads the manifold affine connections $\Gamma^a_{bc}$ to become identical to the Christoffel 
symbols $\{^{\;a}_{bc}\}$ of the metric (i.e., $\Gamma^a_{bc}\equiv\{^{\;a}_{bc}\}$). Hence, a Riemann configuration of spacetime is characterized by the  requirement that $g_{ab\|c}=0$, where the double bar denotes covariant differentiation defined through the Christoffel symbols of the metric [instead of the affine connections $\Gamma^a_{bc}$]. This requirement implies that vector lengths do not change under parallel transport, meaning that the units of measure of the geometry are point-independent. In particular, general relativity with an extra scalar [dilaton] field with the canonical action

$$
S=\int d^{4}x \sqrt{-g}(R-\alpha (\nabla \psi )^2+16\pi L_{matter}),
\eqno{(5)}
$$
where $R$ is the Ricci scalar in terms of the metric $g_{ab}$, $\psi$ is the dilaton field, and $\alpha$ is a free [constant] parameter, is naturally linked with Riemann geometry with point-independent units of measure. It should be noted that, when we put in Eq. (5) $\alpha=0$ [$\psi$ arbitrary] or when $\psi=const$, we recover usual Einstein's formulation of general relativity.

Under the conformal rescaling (1) with $\Omega
^{2}=e^{-\psi }$, the action (5) is mapped into its conformal action 

$$
S=\int d^{4}x \sqrt{-\hat g}(\hat \phi \hat R-\frac{\alpha-\frac{3}{2}}{
\hat \phi}(\hat \nabla \hat \phi)^2+16\pi \hat \phi^2 L_{matter}),  
\eqno{(6)}
$$
where $\hat R$ is the curvature scalar in terms of the metric $\hat g_{ab}$, $\hat\phi$ is the transformed dilaton [the following replacement has been used $\hat\phi=e^{\psi}$], and the matter fields are now non-minimally coupled to the dilaton. At the same time, under this transformation of the spacetime metric, manifolds of Riemann structure are mapped onto manifolds of conformally-Riemannian structure. Therefore, in the theory that is derivable from the action Eq. (6) (Conformal general relativity) the underlying manifolds are of conformally-Riemannian nature. Conformally-Riemannian manifolds are also acknowledged as Weyl-integrable spaces\cite{novello}.\footnotemark\footnotetext{These represent a special case of Weyl geometry that is free of the "second clock effect". This effect is in desagreement with observations\cite{vp}.} In effect, under the rescaling (1) with $\Omega
^{2}=e^{-\psi }$, the Riemannian requirement Eq. (4) is transformed into the following requirement

$$
\hat g_{ab;c}=\psi_{,c}\;\hat g_{ab},
\eqno{(7)}
$$
where now semicolon denotes covariant differentiation in a general affine sense with $\hat\Gamma^a_{bc}$ being the affine connection of the conformal manifold. It is given through the Christoffel symbols of the metric with a hat $\{^{\;a}_{bc}\}_{hat}$ and the derivatives of the scalar function [dilaton] $\psi$ as

$$
\hat\Gamma^a_{bc}\equiv\{^{\;a}_{bc}\}_{hat}-\frac{1}{2}(\psi_{,b}\;\delta^a_c+\psi_{,c}\;\delta^a_b-\hat g_{bc}\hat g^{an}\;\psi_{,n}).
\eqno{(8)}
$$

If one compares Eq. (7) with the requirement of nonvanishing covariant derivative of the metric tensor $\hat g_{ab}$ in the most generic cases of Weyl geometries\cite{novello}:

$$
\hat g_{ab;c}=f_c\;\hat g_{ab},
\eqno{(9)}
$$
in which $f_a(x)$ is the Weyl gauge vector, one arrives at the conclusion that the conformally-Riemannian geometry -characterized by Eq. (7)- is a particular type of Weyl geometry in which the gauge vector is the gradient of a scalar function $\psi$ [the dilaton]. This particular type of Weyl spaces is called a Weyl-integrable spacetime (WIST) since length variations are integrable along closed paths: $\oint dl=0$, where $dl=l\;dx^n\psi_{,n}$, and $l\equiv\hat g_{nm} V^n V^m$ is the length of the vector $V^a(x)$ being parallelly transported along of the closed path. For this reason, in manifolds of WIST configuration, the disagreement with observations due to the "second clock effect"\cite{vp} -that is inherent to Weyl spacetimes in general- is overcome\cite{novello}. 

Therefore, under the transformation (1) with $\Omega^2=e^{-\psi }$ Riemann geometry is mapped into a WIST geometry. In other words, in theories showing non-minimal coupling of the matter fields to the metric, in particular with the matter part of the action of the kind

$$
S_{matter}=16\pi\int d^4x\sqrt{-\hat g}\;e^{2\psi} L_{matter}=16\pi\int d^4x\sqrt{-\hat g}\;\hat \phi^2 L_{matter}
\eqno{(10)}
$$
[it is conformal to (2)], the nature of the underlying manifold is that of a WIST configuration. The equations of free-motion of a material test particle that are derivable from Eq. (10)

$$
\frac{d^2 x^a}{d\hat s^2}+\{^{\;\;a}_{mn}\}_{hat}\frac{dx^m}{d\hat s}\frac{dx^n}{d\hat s}-\frac{\psi_{,n}}{2}(\frac{dx^n}{d\hat s}\frac{dx^a}{d\hat s}-\hat g^{na})=0,
\eqno{(11)}
$$
are conformal to Eq. (3). Equations (11) coincide with the equations defining geodesic curves in spacetimes of WIST configuration. These can also be obtained with the help of the variational principle $\delta\int e^{-\frac{\psi}{2}} d\hat s=0$, that is conformal to $\delta\int ds=0$.

The requirement of nonvanishing covariant derivative of the metric tensor $\hat g_{ab}$ [Eq. (7)] implies that vector lengths may vary along transport or, in other words, that the units of measure may change locally. Therefore WIST geometry represents a generalization of Riemann geometry to include units of measure with point-dependent length.

In other words, under the conformal rescaling (1) with $\Omega^2=e^{-\psi }$  theories with minimal coupling of the matter fields to the metric [for instance those with the matter part of the action of the kind (2)] being uniquely linked with manifolds of Riemann structure, are mapped into theories with non-minimal coupling of the matter fields to the metric [with the matter part of the action of the kind (10)] in which the nature of the underlying manifold is of WIST structure. In particular, the Riemannian geodesics of the metric $g_{ab}$ are mapped under this transformation onto geodesics in manifolds of a WIST structure that are specified by the conformal metric $\hat g_{ab}$ and the gauge vector $\psi_{,a}$ [or, alternativelly, $\hat\phi_{,a}/\hat\phi$]. This means that under (1) the metric $g_{ab}$ that can be measured with the help of clocks and rods made of ordinary matter in manifolds of Riemann geometry is mapped into the metric $\hat g_{ab}$ that can be measured with the help of clocks and rods made of ordinary matter in a manifold of WIST structure. It should be remarked, however, that the metric without a hat and the one with it are not physically equivalent.\footnotemark\footnotetext{Both metrics are however observationally equivalent} In fact, it can be checked that under the transformations of the one-parameter group of conformal transformations\cite{far}

$$
\hat g_{ab}=\hat\phi^{-\sigma} g_{ab},\;\;\; 
\hat\phi=\phi^{1-\sigma},\;\;\;
\hat\alpha=\frac{\alpha}{(1-\sigma)^2}, 
\eqno{(12)}
$$
$\sigma\neq 1$, the action (6) is invariant in form while the action (5) is not invariant under the transformations of this group. In Ref. \cite{iq} it has been argued that this group of transformations can be identified with the one-parameter group of point-dependent transformations of the units of length, time, and mass.\footnotemark\footnotetext{The transformation (1.1) with $\Omega^2=e^{-\psi }=\hat\phi^{-1}$ does not belong to this group since it is the particular case when in Eq. (12) $\sigma=1$} Hence, the action (6) has a higher degree of symmetry than its conformal action (5): It is invariant under point-dependent transformations of the units of length, time, and mass. 

In this letter we shall study an idea that is based on the de Broglie-Bohm quantum theory of motion [first presented in reference \cite{sho}] within the frame of WIST geometry being naturally linked with the conformal
formulation of general relativity [action(6)]. In this context we shall briefly discuss on the Mach's principle and we shall hint at its possible connection with the quantum.

Recently it was shown \cite{sho} that, within the context of the de
Broglie-Bohm relativistic quantum theory \cite{boh}, the quantum effects of
matter can be explicitly included in the metric through a conformal
transformation of the metric. In fact, under a conformal transformation of
the kind (1), the geodesic equations of Riemann geometry [Eq.(3)] are
mapped into the Eq.(11) defining a non-geodesic motion on a spacetime with metric $\hat g_{ab}$ provided that the Riemaniann structure of the geometry is preserved under (1).

If we set $\hat\phi=1+Q$, where $Q$ is the matter quantum potential 
\cite{boh}, hence the last term [third term] in the left hand side (LHS) of
Eq.(11) represents the quantum force \cite{sho}. This quantum force causes a desviation of the motion of a particle from being geodesic [in terms the metric field $g_{ab}$] to being a non-geodesic one [in terms the metric field $\hat g_{ab}$]. We recall that a basic assumption here is that the Riemaniann nature of the geometry is preserved under (1). We are not concerned in this letter with the validity of the ideas developed in Ref.\cite{sho}. Our goal here is to study the consequences of taking an action of the kind (6) for the description of the laws of gravity [the underlying manifold is of WIST structure] when the de Broglie-Bohm quantum theory of motion is approached. In this context, following Ref.\cite{sho}, we shall set $\hat\phi\equiv 1+Q$, where, as before, $Q$ is the matter quantum potential.

The basic consideration in this letter is that, under (1), spacetimes of Riemannian structure are mapped onto Weyl-integrable spacetimes. Correspondingly, the equation (3) defining geodesic curves in Riemann geometry is mapped into the equation (11) defining geodesic curves in spacetimes of WIST configuration. This means that a free-falling test particle would not 'feel' the quantum force if the motion is interpreted on the grounds of a WIST geometry and, correspondingly, the metric with a hat $\hat g_{ab}$ is taken to give the physical interpretantion of the experimental observations. This hints at the conclusion that Weyl-integrable geometry contains implicitly the quantum effects of matter. Consequently the conformal formulation of general relativity [action (6)] -being naturally linked with spacetimes of WIST structure- is already a theory of gravity implicitly containing the quantum effects. Hence, we feel, it is not causal that under (1) incomplete
geodesics over Riemann geometry can be mapped [in principle] into complete
geodesics over a WIST geometry\cite{qbc,iq}. This last geometry is compatible with singularity free spacetimes
even if its conformal [Riemann] geometry is linked with spacetimes
containing singularities. We recall that the inevitabillity of spacetime
singularities in spacetimes of Riemann geometry is usually linked with the
lack of quatum cosiderations in the canonical formulation of general
relativity [action (5)].

The non-local character of the matter quantum potential $Q$ allows one to
consider in a natural way the validity of the Mach's principle within the
context of conformal general relativity and, correspondingly,
within the context of a WIST geometry. Under the rescaling (1)
the inertial mass of a given elemetary particle transforms like \cite{qbc,iq}
 
$$
\hat m=e^{-\frac{1}{2}\psi}m  
\eqno{(13)}
$$
where $m$ is the constant mass of the particle in canonical general
relativity. Correspondingly $\hat m$ is the variable inertial mass of
the particle within the context of conformal general relativity . We recall
that the geometrical interpretention of the experimental observations is to
be given on spacetimes of WIST structure [with the law (7) of lenth
transport]. It is very encouraging that the mass of a single particle, being local in nature, is determined by a global parameter such as the curvature $k$ that characterizes the entire universe. We shall show this with an example. For simplicity let us take an homogeneous and isotropic dust-filled Friedmann-Robertson-Walker (FRW) universe with the Riemannian line-element 

$$
ds^2=-dt^2+a^2 (\frac{dr^2}{1-kr^2}+r^2 d\Omega^2),  
\eqno{(14)}
$$
where $a$ is the time-dependent Riemannian scale factor, $k=0,\pm 1$ is the spatial curvature and $d\Omega^2=d\theta^2+sin^2 \theta d\varphi^2$. The field equations derivable from the action (5) for canonical general
relativity are the following.

$$
(\frac{\stackrel{.}{a}}{a})^{2}+\frac{k}{a^{2}}=\frac{8\pi }{3}\mu +\frac{
\alpha }{6}\psi ^{2},\;\;\; 
\stackrel{..}{\psi }=-3\frac{\stackrel{.}{a}}{a}\stackrel{.}{\psi },\;\;\;
\stackrel{.}{\mu }+3\frac{\stackrel{.}{a}}{a}\mu =0, 
\eqno{(15)}
$$
where the overdot means derivative with respect to the Riemannian cosmic
time $t$. The last equation in (15) can be integrated to give $\mu =\frac{3}{8\pi}\frac{M}{a^3}$ [$M$ is an integration constant]. 

After integrating once the second equation in (15) one obtains

$$
\stackrel{.}{\psi}=\pm\frac{N}{a^3},  
\eqno{(16)}
$$
where $N$ is another integration constant and the "+" and "-" signs define two possible branches of our solution. For further simplification of the analysis we shall consider the particular case with $\alpha =0$ in the first equation in (15).\footnotemark\footnotetext{This value for $\alpha$ is supported by current experiments confirming Einstein's general relativity} Hence, after the change of time variable $dt=\frac{a^2}{\sqrt{M}}d\eta$ this last equation can be readily integrated to give

$$
a=\frac{4}{\eta ^{2}+\frac{4k}{M}}.  
\eqno{(17)}
$$

Consequently integration of Eq.(16) leads to the following
result

$$
\psi^\pm=\psi_{0}\pm \frac{N}{12\sqrt{M}}\eta (\eta^2+\frac{12k}{M}),
\eqno{(18)}
$$
where $\psi _{0}$ is yet another integration constant. Without lost of
generality we can take it to be zero. Hence, if we
substitute (18) in Eq.(13) we see that the inertial mass of an
elementary particle [being a local feature of matter] is affected by the
integration constants -given through the boundary conditions- and by the curvature of the universe ($k$). If we take into account the Eq.(18) then the matter quantum potential can be implicitly written in terms of the integration constants and the curvature $k$ through [$e^\psi=1+Q$]

$$
Q^{\pm}=\exp[\pm \frac{N}{12\sqrt{M}}\eta (\eta ^{2}+\frac{12k}{M})]-1.
\eqno{(19)}
$$

Take, for simplicitly, a flat FRW universe [$k=0$]. In this case the time
coordinate $\eta $ takes values in the interval $0\leq \eta \leq \infty $
[the case $-\infty \leq \eta \leq 0$ represents the time-reversed
situation]. Hence, since the Newton's constant is proportional to $e^{-\psi }$, we shall choose the "-" branch of the solution (18) as the "physical"
branch. In fact, in this branch of the solution the Newton's constant
evolves from being the unity at $\eta =0$ [the scale factor $a$ is infinite] to being infinite in the infinite future $\eta =+\infty $ [the Riemaniann scale factor $a$ is zero meaning that, within the context of Riemann geometry, the universe collapses into a final singularity] as it should be.

As we have properly remarked, the physical interpretation of our results
is to be given in terms of the magnitudes with a hat and, correspondingly within the context of a Weyl-integrable geometry. In the "-"
branch of the solution, the conformal cosmic time $\hat t$ is given in
terms of the time coordinate $\eta $ through the following approximate
relationship $\hat t \sim \int \frac{e^{3\eta}}{\eta^4}d\eta$ while
the conformal scale factor $\hat a \sim \frac{e^{3\eta }}{\eta ^2}$.
It can be verified that, in this case, the flat FRW universe, when
interpreted within the context of a WIST geometry [magnitudes
with a hat], is free of the cosmological singularity\cite{qbc,iq}.
It evolves from being infinite in the infinite past [$\hat t=-\infty$]
trhough a bounce at a minimum value of the scale factor $\hat a$ to
being infinite in the infinite future [$\hat t=+\infty$]. This
bouncing universe is regular everywhere. The matter quantum potential
monotonically decreases from $Q=0$ at the infinite past to $Q=-1$ at the
infinite future. Hence, since the absolute value of the quantum potential increases with the evolution of the universe, this means that, in this particular case, the contribution from the quantum increases with the evolution of the universe. I. e., the universe evolves from being classical in the infinite past to being quantum in nature in the infinite future. Just the opossite situation is obtained if in the untransformed [Riemannian] description the universe evolves from a global cosmological singularity at some initial time [the time-reversed situation].

Summing up. If we study the de Broglie-Bohm approach to the quantum description of the world within the context of the conformal formulation of general relativity [that is naturally linked with a Weyl-integrable geometry], two main conclusions raise. First, the quantum effects of matter are implicitly contained in the Lagrangian formulation of the theory through the dilaton field. In other words, the quantum effects are already implicit in the laws of the Weyl-integrable geometry. The second conclusion is linked with the fact that the Mach's principle, being quantum in nature, acts at the scale of the entire universe, i.e., it acts like a bridge between the micro and the macro scales.

We acknowledge many colleagues for their interest in the ideas
developed in this paper. We, also, thank MES of Cuba by financial support.

\end{document}